\documentclass[aps,prl,showpacs,amsmath,superscriptaddress,preprint,floatfix,nobalancelastpage]{revtex4-1}

\usepackage[nottoc,numbib]{tocbibind}

\usepackage[colorlinks=true, citecolor=black, linkcolor=black, urlcolor=black]{hyperref}

\usepackage[all]{hypcap}
\usepackage{amsmath}
\usepackage{enumerate,bm}
\usepackage{graphicx}
\usepackage{grffile} 
\usepackage{color}
\usepackage{esint}
\usepackage{textpos}
\usepackage[usenames,dvipsnames]{xcolor}
\usepackage{subfigure}
\usepackage{flushend}
\usepackage{tabularx}
\usepackage{amssymb}
\usepackage{float}
\usepackage{subfigure}
\usepackage{ulem} 

\usepackage{setspace}
\usepackage{verbatim}

\begin{document}

\author{Simone~Latini}
\email{simone.latini@mpsd.mpg.de}
\affiliation{Max Planck Institute for the Structure and Dynamics of Matter and Center for Free Electron Laser Science, 22761 Hamburg, Germany}

\author{Dongbin~Shin}
\affiliation{Max Planck Institute for the Structure and Dynamics of Matter and Center for Free Electron Laser Science, 22761 Hamburg, Germany}

\author{Shunsuke~A.~Sato}
\affiliation 
{Center for Computational Sciences, University of Tsukuba, Tsukuba 305-8577, Japan}
\affiliation{Max Planck Institute for the Structure and Dynamics of Matter and Center for Free Electron Laser Science, 22761 Hamburg, Germany}

\author{Christian~Sch\"afer}
\affiliation{Max Planck Institute for the Structure and Dynamics of Matter and Center for Free Electron Laser Science, 22761 Hamburg, Germany}

\author{Umberto~De~Giovannini}
\affiliation{Max Planck Institute for the Structure and Dynamics of Matter and Center for Free Electron Laser Science, 22761 Hamburg, Germany}
\affiliation{Nano-Bio Spectroscopy Group,  Departamento de Fisica de Materiales, Universidad del País Vasco UPV/EHU- 20018 San Sebastián, Spain}
\affiliation{Nano-Bio Spectroscopy Group,  Departamento de Fisica de Materiales, Universidad del País Vasco UPV/EHU- 20018 San Sebastián, Spain}

\author{Hannes~H\"ubener}
\affiliation{Max Planck Institute for the Structure and Dynamics of Matter and Center for Free Electron Laser Science, 22761 Hamburg, Germany}

\author{Angel~Rubio}
\email{angel.rubio@mpsd.mpg.de}
\affiliation{Max Planck Institute for the Structure and Dynamics of Matter and Center for Free Electron Laser Science, 22761 Hamburg, Germany}
\affiliation{Nano-Bio Spectroscopy Group,  Departamento de Fisica de Materiales, Universidad del País Vasco UPV/EHU- 20018 San Sebastián, Spain}
\affiliation{Center for Computational Quantum Physics (CCQ), The Flatiron Institute, 162 Fifth avenue, New York NY 10010.}

\title{The Ferroelectric Photo-Groundstate of SrTiO$_3$: Cavity Materials Engineering}

\date{\today}

\maketitle 

{\bf Optical cavities confine light on a small region in space which can result in a strong coupling of light with materials inside the cavity. This gives rise to new states where quantum fluctuations of light and matter can alter the properties of the material altogether.
Here we demonstrate, based on first principles calculations, that such light-matter coupling induces a change of the collective phase from quantum paraelectric to ferroelectric in the SrTiO$_3$ groundstate, which has thus far only been achieved in out-of-equilibrium strongly excited conditions~\cite{Li2019,Nova2019}. This is a light-matter-hybrid groundstate which can only exist because of the coupling to the vacuum fluctuations of light, a \textit{photo-groundstate}. The phase transition is accompanied by changes in the crystal structure, showing that fundamental groundstate properties of materials can be controlled via strong light-matter coupling.
Such a control of quantum states enables the tailoring of materials properties or even the design of novel materials purely by exposing them to confined light.}

\section*{Introduction}
Engineering an out of equilibrium state of a material by means of strong light fields can drastically change its properties and even induce new phases altogether. This is considered a new paradigm of material design, especially when the collective behaviour of particles in quantum materials can be controlled to provide novel functionalities~\cite{Tokura:2017bh, Hsieh:2017ix}.
Alternative to the intense lasers necessary to reach such out of equilibrium states, one can achieve strong light-matter coupling by placing the material inside an optical cavity~\cite{liu2015, li2018, ruggenthaler2018, schafer2018ab, paravicini2019, delteil2019, lupatini2020, Hubener:2020fm}. A main advantage of this approach is that strong interaction can be achieved at equilibrium, opening up new possibilities for materials manipulation. Among the proposed effects are novel exciton insulator states~\cite{Mazza2019}, control of excitonic energy ordering~\cite{Latini:2019}, enhanced electron-phonon coupling~\cite{Sentef:2018gp}, photon mediated electron pairing~\cite{Schlawin:2019jw, curtis2019, allocca2019, raines2020cavity} and enhanced ferroelectricity~\cite{Ashida2020}.
One enticing possibility is, however, to change the ground state of a material and to create a new phase not through excited quasi-particles but truly as the equilibrium state. 

Here we show that this can be achieved in the paraelectric SrTiO$_3$ as a photo correlated ferroelectric groundstate. This groundstate, which we refer to as \textit{photo-groundstate}, is the result of the strong coupling between matter and quantum vacuum fluctuations of light. While similar materials of the perovskite family undergo a para- to ferroelectric phase transition at low temperatures, SrTiO$_3$ remains paraelectric~\cite{Song1996}, because the nuclear quantum fluctuations prevent the emergence of a collective polarization that is characteristic of the ferroelectric phase~\cite{Muller:1979je, Zhong1996}. Alterations to the material that overcome a relatively small activation energy, however, can induce ferroelectricity: for instance through isotope substitution~\cite{Itoh:1999eq}, strain~\cite{Li:2006, Haeni2004} and most notably nonlinear excitation of the lattice by strong and resonant terahertz laser pumping~\cite{Li2019,Nova2019}. In the latter type of experiments a transient broken symmetry of the structure as well as macroscopic polarization indicative of a transient ferroelectric phase have been observed.

By using atomistic calculations we show that the off-resonant dressing of the lattice of SrTiO$_3$ with the vacuum fluctuations of the photons in a cavity can suppress the nuclear quantum fluctuations in a process that is analogous to the one of dynamical localisation~\cite{Dunlap:1986co}: the interaction with cavity photons effectively results in an enhancement of the effective mass of the ions thus slowing them down and reducing the importance of their quantum fluctuations. We further demonstrate that the effect of cavity induced localization extends to finite temperatures, even when thermal lattice fluctuations overcome the quantum ones. We thus introduce a revisited paraelectric to ferroelectric phase diagram, with the cavity coupling strength as new dimension.

\section*{Theory} 
\begin{figure}[t]
  \centering
   \includegraphics[width=\columnwidth]{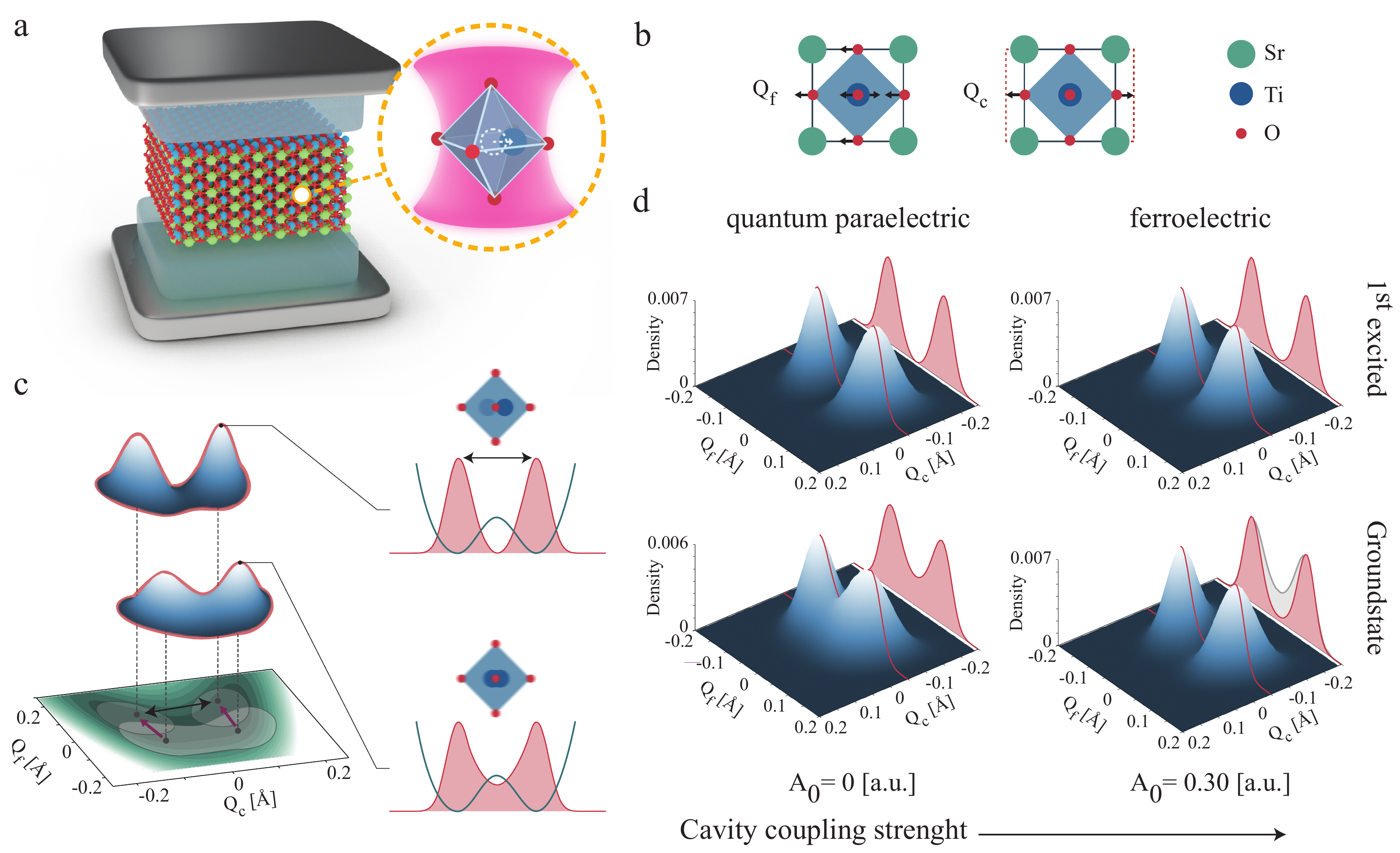}
   \caption{\label{fig:1}
   {\bf{The emergence of ferroelectricity in SrTiO$_3$ in a dark optical cavity}}. (a) Cartoon of SrTiO$_3$ embedded in a dielectric medium inside a Fabry-Perot cavity. The zoomed sketch shows the direction of the Ti-O displacive motion induced by the photons in the cavity. (b) Illustration of the main lattice motions involved in ferroelectricity: the ferroelectric mode and lattice vibration along the same direction, parameterized with $Q_{\rm{f}}$ and $Q_{\rm{c}}$ respectively. (c) Contour plot of the first principles PBE 2D potential energy surface and a schematic representation of the nuclear groundstate localization due to the coupling of the ferroelectric mode to the photons in the cavity. (d) First principles orbital resolved density associated to the ground and first excited states outside the cavity ($A_0=0$~a.u.) and inside the cavity ($A_0=0.3$~a.u.) for a cavity frequency of $3$~THz.}
\end{figure}

A microscopic theory that describes the structural and polarization properties of SrTiO$_3$ has to inherently include nuclear quantum fluctuations, as confirmed by extensive literature~\cite{Muller:1979je, Zhong1996, Song1996}. Ref.~\cite{dongbin2021} develops a first principles approach that includes the quantum mechanical nature of the ions and finds that the phases of SrTiO$_3$ can be properly described by specifically accounting for the quantum fluctuations of the non-linearly coupled ferroelectric soft (FES) mode and lattice vibration, see sketch in Fig.~\ref{fig:1}(b). In SrTiO$_3$ the quantum fluctuations of the two vibrational modes are strong enough to wash out the localization imposed by the double well shaped potential energy surface (see Fig.~\ref{fig:1}(c)) and destroy the ferroelectric order stabilizing the so called quantum paraelectric phase ~\cite{Muller:1979je, Zhong1996}. 
In this work we introduce the paradigm of altering the localization and hence the macroscopic polarization properties of SrTiO$_3$ by coupling the FES mode to the confined quantized light modes of an optical cavity. The possible setup we consider in the following, is a bulk film of tetragonal SrTiO$_3$ encapsulated in a transparent dielectric embedded in a Fabry-Pérot cavity as sketched in Fig.~\ref{fig:1}(a). Here we choose the SrTiO$_3$ crystal c-axis to be parallel to the cavity mirrors and consider a single cavity photon mode with the smallest allowed momentum along the perpendicular direction to the mirrors. Because the displacement of the Ti-O atoms creates a dipole, the FES mode couples to the electric field of the confined photons. Despite the specific choice of the cavity geometry for the results presented in the following, the same ideas can be extended to other configurations or materials. The setup described above can be cast into an atomistic quantum electrodynamical (QED) Hamiltonian for the single unit cell which reads: $\hat{H} = \omega_{\rm c} \hat{a}^{\dagger}\hat{a} + \frac{\hat{p}_{\rm{c}}^2}{2 M_{\rm{c}}}  + \frac{1}{2 M_{\rm{f}}}\left[\hat{p}_{\rm f} - A_0 Z_{\rm f}\left(\hat{a}^{\dagger}+\hat{a}\right)\right]^2 + V_{\rm DFT}(\hat{Q}_c,\hat{Q}_{\rm f})$. This Hamiltonian describes the coupling of the zero-momentum FES mode with a single cavity photon mode of frequency $\omega_{\rm c}$ and with a coupling strength determined by $A_0Z_{\rm f}/M_{\rm f}$ with $A_0$ the effective mode volume, $Z_{\rm f}$ the effective charge and $M_{\rm f}$ the effective mass of the FES mode. For details on all the quantities see SM.
Our Hamiltonian builds upon the one reported in Ref.~\cite{dongbin2021} and properly describes the quantum paraelectric phase and the temperature dependence of the FES mode by including the fundamental phonon non-linearities of SrTiO$_3$ via the potential $V_{\rm DFT}$ calculated by density functional theory (DFT).

\section*{Results and Discussion} 
The groundstate and excited states of SrTiO$_3$ dressed with the quantized cavity photons can then be accessed via exact diagonalization of the QED Hamiltonian.
The density corresponding to the matter component of ground and first excited state for material with and without coupling to the cavity photons are calculated by tracing out the photonic part and are reported in Fig.~\ref{fig:1}(d). For SrTiO$_3$ outside the cavity, the groundstate is different from the first excited state of the FES mode, which happens to be the first excited state of the QED Hamiltonian. It is a characteristic of a quantum paraelectric that despite the double well shaped potential energy surface, the quantum fluctuations prevent the system to localize in the wells. When the coupling to the cavity is turned on, the groundstate and the first excited states become degenerate, indistinguishable and present a clear node at $Q_{\rm f}=0$. This is indicative of a ferroelectric phase: the system can choose to be in a linear combination of the two states where the Ti and O atoms are have a finite positive or negative displacement. Any real system will then undergo spontaneous symmetry breaking and localize the FES mode in one of the two wells leading to the formation of a macroscopic polarization, typical of the ferroelectric phase. 
 
To summarize, by coupling the SrTiO$_3$ phonons to the cavity photons we have realized the transition to a ferroelectric groundstate. This phase transition can be explained in terms of the dynamical localization effects~\cite{Dunlap:1986co,Sentef:2020ho}, where the cavity dresses the masses of the lattice modes and thereby reduces their quantum fluctuations as depicted in Fig.~\ref{fig:1}{c}. Reducing the dimensionality of the system to the FES mode only and assuming a left localized and a right localized basis the QED Hamiltonian can be simplified to: $H = \omega_c a^\dagger a + t(\hat{c}^{\dagger}_{\rm R}\hat{c}_{\rm L} + \hat{c}^{\dagger}_{\rm L}\hat{c}_{\rm R})  - iA_0 \frac{Z_{\rm f}}{M_{\rm f}} [(\hat{c}^{\dagger}_{\rm R}\hat{c}_{\rm L}-\hat{c}^{\dagger}_{\rm L}\hat{c}_{\rm R})(a+a^\dagger)]$, where $\hat{c}^{\dagger}_{\rm L/M}$ and $\hat{c}_{\rm L/M}$ are the creation and annihilation operator of the left and right states. If the hopping value $t$ is chosen to be big enough to overcome the double well potential barrier and it results in quantum paraelectricity. In the presence of the cavity photons, a simple effective Hamiltonian can be derived, as shown in the SM, and in the high cavity frequency limit reads $H_{\rm eff} = \left[t - \left(\frac{A_0 Z_{\rm f}}{M_{\rm f}\omega_c}\right)^2 t\right](\hat{c}^{\dagger}_{\rm R}\hat{c}_{\rm L} + \hat{c}^{\dagger}_{\rm L}\hat{c}_{\rm R}) - \frac{A_0^2 Z_{\rm f}^2 }{M_{\rm f}\omega_c}(\hat{c}^{\dagger}_{\rm R}\hat{c}_{\rm R}+\hat{c}^{\dagger}_{\rm L}\hat{c}_{\rm L})$. Hence the effect of the photon cavity is to localize the system by effectively reducing the hopping between the left and right states, $t \rightarrow t\left[1 -\left(\frac{A_0 Z_{\rm f}}{M_{\rm f}\omega_c}\right)^2\right]$. 

\begin{figure}[t]
  \centering
   \includegraphics[width=\columnwidth]{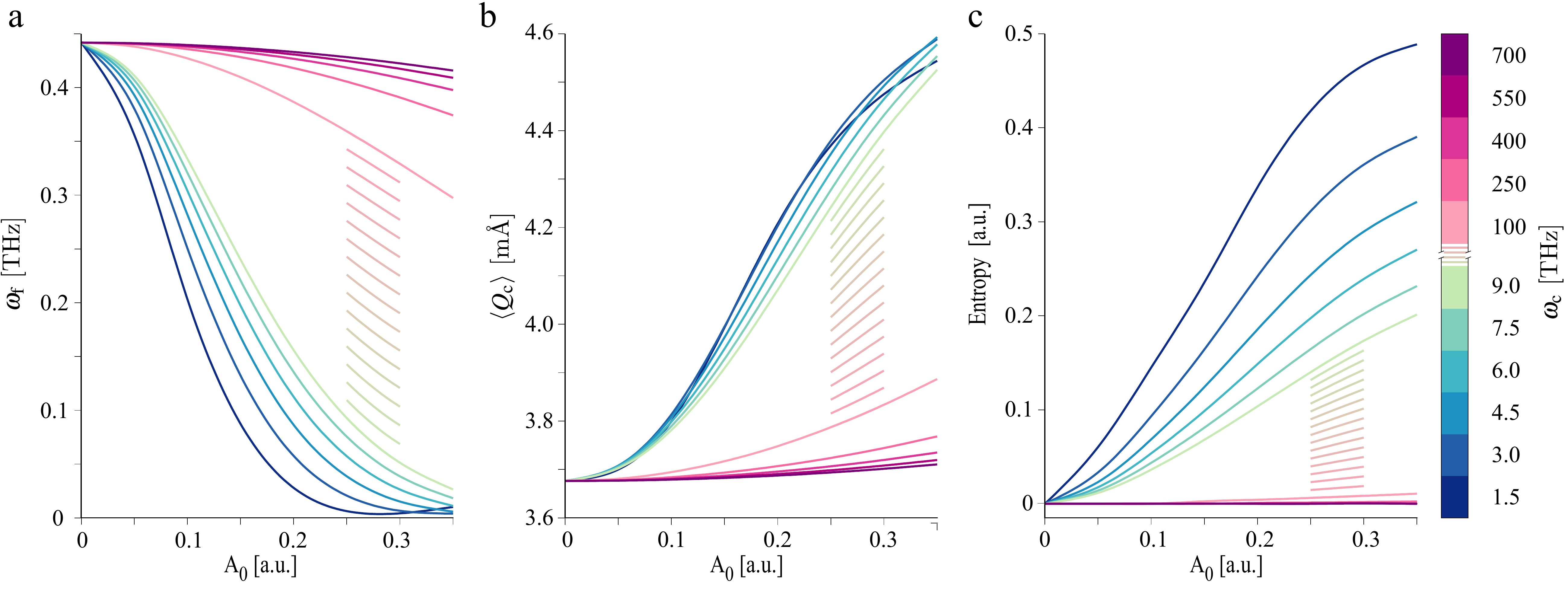}
   \caption{\label{fig:2}
   {\bf Dependence of the microscopic properties of SrTiO$_3$ on the cavity frequency $\omega_c$ and coupling strength $A_0$ expressed in atomic units (a.u.).} (a) Generalized ferroelectric soft-mode frequency, calculated as the difference in the energy of the nuclear first excited and ground state as a function of the cavity coupling strength for different cavity photon energies. Expectation value of the (b) mean displacement of the $c$ lattice parameter as a function of the cavity coupling and photon energy. (c) Von-Neumann entropy of the photonic subsystem as a function of cavity coupling and photon energy. This quantity indicates the degree of correlation going beyond mean-field (Maxwell like) light-matter coupling.}
\end{figure}

In the following we report on the dependence of the ferroelectric soft mode frequency, lattice displacements and subsystem entropy as function of the cavity coupling strength and the frequency of the cavity at zero temperature. In Fig.~\ref{fig:2}(a) we show the FES mode frequency identified as the energy difference between the first excited state and groundstate of the QED Hamiltonian. This difference is reduced with increasing cavity coupling corresponding to a softening of the FES mode, which is indicative of the transition the ferroelectric phase. 
However, we note that when the system acquires a ferroelectric character, the FES mode is no longer a proper normal mode because of the non negligible bilinear coupling with the lattice vibrations at the bottom of the wells. Therefore we refer to this energy difference as the \textit{generalized} FES mode frequency. The striking result is that ferroelectricity can be reached for a wide range of cavity photon energies, in other words the cavity does not need to be resonant with the generalized FES mode energy (or any other phonon modes) and indeed the effect is larger at off-resonance. This result is of technological relevance since in the recent experiments that report on laser-induced ferroelectricity the laser had to be in resonance with the FES mode at $~0.5$~THz, which is a challenge for laser technology~\cite{Li2019}. Considering the expectation value of the squared ferroelectric displacement, shown in SM, we find that the largest localization of the FES mode is achieved at around $\omega_{\rm c}=3$~THz. The expectation value of the lattice vibration displacement in Fig.~\ref{fig:2}(b) follows a similar trend and for increasing coupling to the cavity photons we can observe a lattice expansion which has important consequences on the FES mode. Indeed, as shown in Fig.~\ref{fig:1}(c), the double well for the expanded lattice deepens and hence the localization of the FES mode is enhanced. This, together with the effect of dynamical localization is the mechanism underlying the transition to ferroelectricity. 
The analysis of the photon component of the groundstate indicates that even though the empty cavity is dark, the light-matter interaction creates a finite photon number. We therefore refer to the groundstate inside the cavity as a \textit{photo-groundstate}. As a measure of light-matter correlation, we evaluated the von-Neumann entropy for the photonic subsystem, c.f. SM. This quantity, reported in Fig.~\ref{fig:2}(d) indicates whether the system can be represented as a simple tensor product of a matter and a photonic state. The entropy becomes non-zero and increases with increasing coupling which in a groundstate can only be the result of vacuum fluctuations. 
While in Ref.~\cite{Ashida2020} it was suggested that the ferroelectric phase of SrTiO$_3$, that is induced by external perturbation, can be enhanced by the cavity-matter coupling, we demonstrated here that the ferroelectric phase can be reached as an unperturbed photo-groundstate, when the intrinsic phonon non-linearities are accounted for.

The cavity coupling strengths and cavity frequencies considered here can be achieved by tuning the SrTiO$_3$ film thickness and the distance between the cavity mirrors. Because the light-matter coupling is proportional to $\sqrt{d_{\perp}/v}$, with $d_{\perp}$ and $v$ the thickness of SrTiO$_3$ and the unit cell volume respectively, a coupling value up to $A_0=0.35$~a.u. (atomic units) can be achieved for a thickness up to $d_{\perp}\sim 20~\mu$m, while a frequency in the range of $1$ to $10$~THz require cavity lengths between $L_{\perp}\sim 300$ to $L_{\perp}\sim 30~\mu$m. A more extensive discussion and the definition of the light-matter coupling is reported in the SM.

\begin{figure}
   \includegraphics[width=1.0\columnwidth]{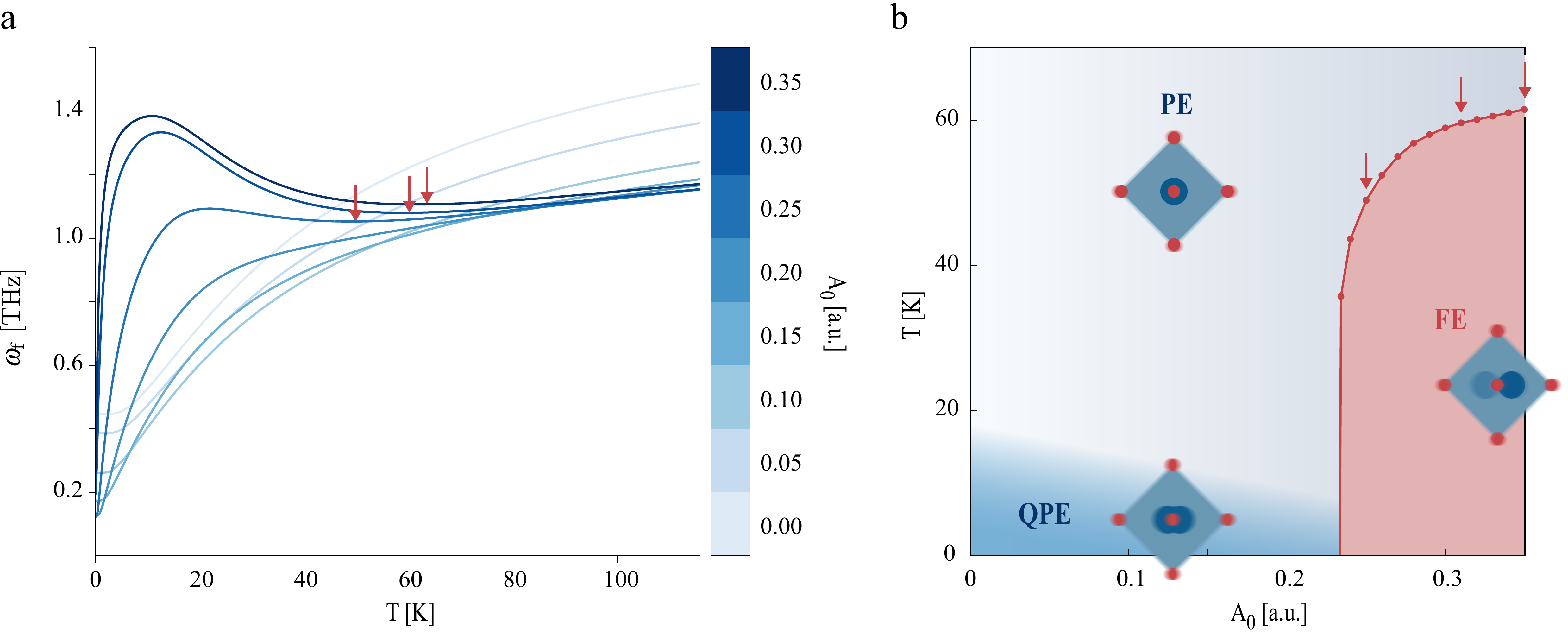}
   \caption{\label{fig:3}
    {\bf Revisited phase diagram of SrTiO$_3$ inside the optical cavity}. (a) Effective microscopic ferroelectric frequency as a function of lattice temperature for different cavity coupling strengths. The minima of the curves, marked with red circles, are used to identify the phase boundaries in panel (b). (b) The microscopic phase diagram of SrTiO$_3$ inside the cavity. Three possible scenarios are opened up by the coupling to confined light. At low temperature increasing the coupling with the cavity leads to a transition from a quantum paraelectric to a ferroelectric; at intermediate temperatures increasing the coupling first allows thermal fluctuations to overcome the quantum one and eventually reach a transition between paraelectric to ferroelectric; at high temperatures, since thermal fluctuations are much higher than the quantum ones a direct transition from para- to ferroelectric can be expected for increasing coupling to the cavity.}
\end{figure}

We now revisit the phase diagram of SrTiO$_3$ embedded in a quantum cavity by adding the tunable coupling strength as a dimension of the diagram. Here we suggest a microscopic approach to estimate a phase diagram by accounting for the effect of temperature via Kubo's linear response theory on a thermal equilibrium state (see SM for details). We stress that such an approach neglects the phonon-phonon interactions and the temperature dependent entropic effects arising from the excitation of higher momentum phonon modes which are not computationally feasible, but it allows us to keep our theory parameter free. The response is calculated with respect to an external electric field coupled to the FES mode representing a probe of ferroelectricity. The calculated response function for a cavity frequency $\omega_{\rm c}=3$~THz is shown in the SM and is characterized by a series of peaks whose intensity and width are changing with cavity coupling strength and temperature. In the case of no cavity coupling the average of the frequency weighted by the response can be identified as the frequency of the FES mode, which displays as a function of temperature, a minimum resulting from a characteristic softening and a subsequent stiffening~\cite{Cowley:1979}. This behavior, shown in Fig.~\ref{fig:3}(a) is a hallmark of the phase transition from paraelectric to ferroelectric and we therefore identify the transition temperature as the position of the minimum of the FES mode frequency. We stress that despite the paraelectric to ferroelectric transition being a macroscopic collective phenomenon that involves the generation of a finite static polarization field, it leaves a signature on a microscopic dynamical quantity, the FES frequency, which can be used to estimate the transition.
We then extend the frequency averaging procedure to finite cavity coupling and show the results in Fig.~\ref{fig:3}(a) where a distinctive trend can be observed for cavity couplings beyond a certain threshold. By tracing the evolution of the minimum characteristic of the paraelectric to ferroelectric phase transition we are able to define a phase-diagram which illustrates how the macroscopic phase of SrTiO$_3$ can be controlled by changing temperature and the cavity coupling strength. In the same phase diagram we have also indicated the low temperature quantum paraelectric phase with a gradient color, because the transition from paraelectric to quantum paraelectric is not an actual phase transition~\cite{Zhong1995}. The difference between a quantum paraelectric and a paralectric can rather been understood in terms of whether thermal fluctuations prevails the quantum ones. In this sense we expect that the temperature at which the thermal fluctuations dominate decreases with increasing coupling strength, because the quantum fluctuations are suppressed by the cavity.
A direct application of the phase-diagram would be to prepare a paraelectric SrTiO$_3$ inside an optical cavity close to the phase boundary so that a laser resonant with cavity photon can induce ferroelectricity. This approach has the advantage that the laser can be weak and not in resonance with the low frequency of the FES mode. This can be seen as enhanced ferroelectricity along the lines of what has been proposed in Ref.~\cite{Ashida2020}.

\section*{Conclusion} 
The concept of photo-groundstate introduced from first principles calculations in this work presents a new paradigm for control of materials properties and opens new avenues for materials engineering~\cite{Hubener:2020fm}. The notable property here is that the vacuum fluctuations of the photon field dress the groundstate and alter the crystal structure, lattice constant and phonon frequency of the materials and even stabilize a macroscopic phase. Similar manipulations in other materials can be envisaged to yield control over quantum properties such as  magnetic or even superconducting states.

\section*{Acknowledgements}
We are grateful for the illuminating discussions with Dmitri Basov, Atac Imamoglu and Jerome Faist and Jean-Marc Triscone, Peter Littlewood, Andrew Millis, Michael Ruggenthaler, Michael A. Sentef and Eugene Demler. We acknowledge financial support from the European Research Council(ERC-2015-AdG-694097). Grupos Consolidados (IT1249-19), JSPS KAKENHI Grant Number JP20K14382 and the Cluster of Excellence 'CUI: Advanced Imaging of Matter' of the Deutsche Forschungsgemeinschaft (DFG) - EXC 2056 - project ID 390715994. The Flatiron Institute is a division of the Simons Foundation. S. L. and D. S. acknowledge support from the Alexander von Humboldt foundation.

\end{document}


\title{Supplementary Materials:\\
The Ferroelectric Photo-Groundstate of SrTiO$_3$ : Cavity Materials Engineering}

\author{Simone~Latini}
\email{simone.latini@mpsd.mpg.de}
\affiliation{Max Planck Institute for the Structure and Dynamics of Matter and Center for Free Electron Laser Science, 22761 Hamburg, Germany}

\author{Dongbin~Shin}
\affiliation{Max Planck Institute for the Structure and Dynamics of Matter and Center for Free Electron Laser Science, 22761 Hamburg, Germany}

\author{Shunsuke~A.~Sato}
\affiliation 
{Center for Computational Sciences, University of Tsukuba, Tsukuba 305-8577, Japan}
\affiliation{Max Planck Institute for the Structure and Dynamics of Matter and Center for Free Electron Laser Science, 22761 Hamburg, Germany}

\author{Christian~Sch\"afer}
\affiliation{Max Planck Institute for the Structure and Dynamics of Matter and Center for Free Electron Laser Science, 22761 Hamburg, Germany}

\author{Umberto~De~Giovannini}
\affiliation{Max Planck Institute for the Structure and Dynamics of Matter and Center for Free Electron Laser Science, 22761 Hamburg, Germany}
\affiliation{Nano-Bio Spectroscopy Group,  Departamento de Fisica de Materiales, Universidad del País Vasco UPV/EHU- 20018 San Sebastián, Spain}

\author{Hannes~H\"ubener}
\affiliation{Max Planck Institute for the Structure and Dynamics of Matter and Center for Free Electron Laser Science, 22761 Hamburg, Germany}

\author{Angel~Rubio}
\email{angel.rubio@mpsd.mpg.de}
\affiliation{Max Planck Institute for the Structure and Dynamics of Matter and Center for Free Electron Laser Science, 22761 Hamburg, Germany}
\affiliation{Nano-Bio Spectroscopy Group,  Departamento de Fisica de Materiales, Universidad del País Vasco UPV/EHU- 20018 San Sebastián, Spain}
\affiliation{Center for Computational Quantum Physics (CCQ), The Flatiron Institute, 162 Fifth avenue, New York NY 10010.}

\maketitle

\section{QED Hamiltonian}
In order to predict the properties of SrTiO$_3$ embedded in an optical cavity we a introduce the following atomistic quantum electrodynamical (QED) Hamiltonian as such~\cite{ruggenthaler2018}: 
\begin{equation}
\hat{H} = \omega_{\rm c} \hat{a}^{\dagger}\hat{a} + \frac{\hat{p}_{\rm{c}}^2}{2 M_{\rm{c}}}  + \frac{1}{2 M_{\rm{f}}}\left[\hat{p}_{\rm f} - A_0 Z_{\rm f}\left(\hat{a}^{\dagger}+\hat{a}\right)\right]^2 + V_{\rm DFT}(\hat{Q}_c,\hat{Q}_{\rm f}),
\end{equation}
where $\omega_{\rm c}$ is the frequency of the photons in the cavity which is set by the cavity length $L_{\perp}$, $a^{\dagger}$ and $a$ are the corresponding creation and annihilation operators, $A=A_0(a^{\dagger}+a)$ is the cavity vector potential, $M_{\rm{c}}=111492$~a.u. and $M_{\rm{f}}=194059$~a.u. the effective masses of the lattice vibration (sum of the masses of the atoms in the unit cell) and FES mode respectively, $Z_{\rm{f}}$ the Born effective charge of the FES mode and $V_{\rm DFT}(Q_c,Q_{\rm f})=\sum_{i=1}^6 k_{f,i}\hat{Q}^{2i}_f + \sum_{j=2}^5 k_{c,j}\hat{Q}^{j}_c+\sum_{i=1}^6\sum_{j=1}^5 k_{fc,i,j}\hat{Q}^{2i}_f \hat{Q}^{j}_c $ the potential energy surface shown in Fig.~1(c) of the main text which includes the intrinsic phonon non-linearities of SrTiO$_3$. The FES and lattice modes are parameterized in terms $Q_{\rm f}$ and $Q_{\rm{c}}$ respectively (see next section for more details on $Q_{\rm f}$. The Born effective charges and the 2D potential energy surface expansion coefficients $k$ are determined within DFT using the PBE functional~\cite{Perdew1996} as described in Ref.~\cite{dongbin2021}. Furthermore, we assume the Born-effective charge $Z_f$ to be not affected by the light-matter coupling. We stress that in the Hamiltonian it is essential to take into account the diamagnetic as it guarantees the existence of a groundstate bound from below~\cite{rokaj2018}. 
It is important to note that in the Hamiltonian above we reduced the phononic degrees of freedom of SrTiO$_3$ to the FES mode and lattice vibration only and it is effectively describing a single unit cell. The unit cell Hamitlonian is effectively describing the collective $\Gamma$-phonon modes coupled with the dipole component of the electromagnetic fields. As shown in Ref.~\cite{dongbin2021}, this two modes are sufficient to correctly describes the physics of SrTiO$_3$ by demonstrating quantum paraelectricity and proving the correct behaviour of FES mode for an extended range of temperatures.

The strength of the cavity light-phonon coupling is determined by the Born effective charge and the photon mode amplitude $A_0$. For the $\Gamma$-phonon mode coupled to the dipole component of the electromagnetic field the mode amplitude is given by~\cite{Ashida2020, Sentef:2018gp}:
\begin{equation}
A_0=\sqrt{\frac{d_{\perp}}{2\pi c ~v}}
\end{equation}
with $d_{\perp}$ and $v$ the thickness of the SrTiO$_3$ slab and the volume of the unit cell respectively, $c$ the speed of light and we used the relation $L_{\perp}=\pi c / \omega_{\rm c}$ where $L_{\perp}$ is the vertical dimension of the cavity. In order to obtain the expression above we took into account the difference in the thickness of the cavity and the material, i.e. $V_{\rm c} = V_{{\rm SrTiO}_3}*L_\perp/d_\perp$.  The scaling of the mode volume above with the thickness of the SrTiO$_3$ crystal provides a direct way to increase the coupling of the cavity with the material.

\section{Parameterization of Lattice Mode}
In the atomistic Hamiltonian shown in the previous section we have conveniently decided to parameterize the FES mode with the distance between the Ti and O atoms along the c-axis, as this distance is ultimately involved in the definition of the unit cell dipole: $|\vec{D}| \equiv Z_{\rm f}d_{\rm Ti-O}$. The particular choice however implies that the effective mass $M_{\rm f}$ and Born effective charge $Z_{\rm f}$ have to be rescaled accordingly as compared to the corresponding quantities for the FES phonon mode. This is because, even if of minor importance, the FES phonon mode involves the motion of all the other atoms in the unit cell.
Since within DFTP@PBE~\cite{Gonze1997} the FES mode has an imaginary frequency, the phonon mode is ill defined and therefore we evaluated such mode directly from the difference of the atomic position in the optimized paraeletric and ferroelectric geometry. Specifically, we defined the FES eigenvector as:
 \begin{equation}
\vec{U}^{\rm f}_{I} =  \frac{\vec{s}^{\rm ~ferro}_{I}-\vec{s}^{\rm ~para}_{I}}{\sum_J |\vec{s}^{\rm ~ferro}_{J}-\vec{s}^{\rm ~para}_{J}|},
\end{equation}
where $I$ is the index running over the atoms of the unit cell and $\vec{s}$ are the atomic basis vectors of the two different geometries. The eigevector above and $Q_{\rm f}$ are then related by:
\begin{equation}
    Q_{\rm f} = x_{\rm f} \left( U^{\rm f}_{{\rm Ti},z}-U^{\rm f}_{{\rm O},z}\right)
\end{equation}
with $x_{\rm f}$ the actual FES mode parameter, i.e. the one that is independent of the specific atoms and to which the phonon effective masses and Born effective charges are standardly referred to. To calculate the effective mass and charge to the $Q_f$ parameter, the following formula can be applied:
\begin{equation}
\begin{split}
    M_{\rm f} &= \sum_I M_I \left(\frac{U^{\rm{f}}_{I,z}}{U^{\rm{f}}_{Ti, z} - U^{\rm{f}}_{O, z}}\right)^2\\
    Z_{\rm f} &= \sum_a Z_I \left(\frac{U^{\rm{f}}_{I,z}}{U^{\rm{f}}_{Ti, z}- U^{\rm{f}}_{O, z}}\right),
\end{split}
\end{equation}
where $M_I$ and $Z_I$ are the atomic mass and the Born effective charge for the atom $I$.

Diagonalizing our atomistic Hamiltonian for the SrTiO$_3$ outside the cavity, with the quantities defined above calculated using the PBE functional, gives a FES mode frequency of $0.44$~THz which reproduces well the experimental results~\cite{Shirane1969,Yamanaka2000,Vogt1995}. The full diagonalization of the Hamiltonian is performed on a simple product basis set $|Q_{\rm f}\rangle\otimes|Q_{\rm c}\rangle\otimes|n\rangle$ which consists of a $50\times25$ real space grid for the phononic coordinates $Q_{\rm f}$ and $Q_{\rm c}$ and up to $n=9$ Fock number states as a basis for the photons.

\section{Characterization of the Ferroelectric Photo-Groundstate}
\begin{figure}[b]
  \centering
   \includegraphics[width=\columnwidth]{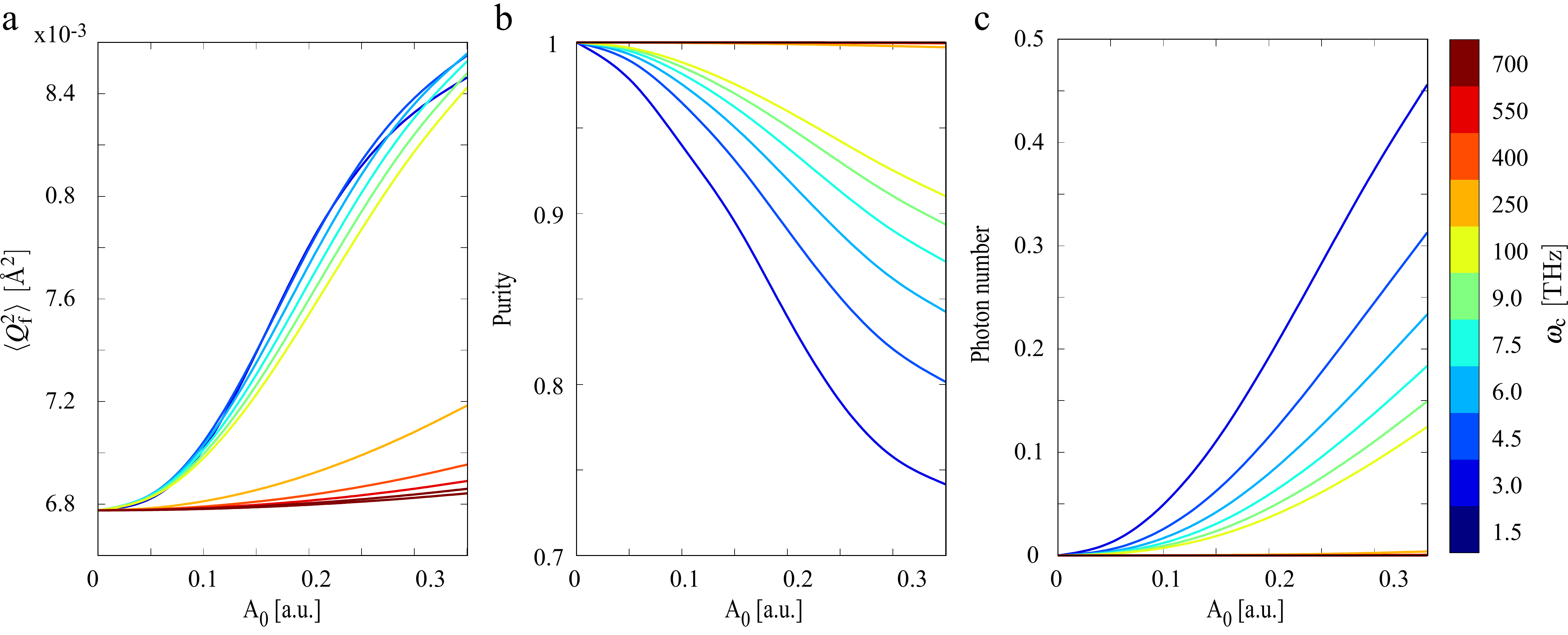}
   \caption{\label{fig:S1}
   {\bf Dependence of the microscopic properties of SrTiO$_3$ on the cavity parameters.} (a) mean squared displacement of the generalized ferroelectric soft mode as a function of the cavity coupling and photon energy. Note that by the symmetry of the 2D potential energy surface the expectation value of the ferroelectric soft mode displacement has to be zero. (b) Purity of the photo-groundstate as a measure of light-matter correlation. (c) Expectation value of the number of photons. }
\end{figure}

In the main text we characterized the photo-groundstate of SrTiO$_3$ in terms of the generalized FES mode frequency, mean displacement of the lattice vibration and Von Neumann entropy of the photonic sub-system. The latter quantity is commonly used to describe the amount of correlation between a given sub-system and all the others, which in our case is just the phononic system. The Von Neumann entropy is defined as:
\begin{equation}
    \mathcal{S} = - \eta_i \sum_{i} \eta_i \log(\eta _i),
\end{equation}
where $\eta_i$ are the eigenvalues of the density matrix of the chosen subsystem, which for the photons is defined as:
\begin{equation}
    \hat{\rho}_{\rm ph} = {\rm Tr}_{\rm pn}\left[\hat{\rho}_{\rm full}\right],
\end{equation}
with ${\rm Tr}_{\rm pn}$ meant as the trace over the phononic states. The resulting photonic entropy is the one shown in Fig.~2(c) of the min text.

To further characterize the photo-groundstate of SrTiO$_3$, we report here the expectation value of the squared FES mode displacement, the purity and the expectation value of the photon number. These quantities are shown in Fig.~\ref{fig:S1} as a function of the coupling strength. 
We point out that the maximum of the mean squared displacement $\langle \hat{Q}_{\rm f}^2\rangle$ is maximum at $\omega_c=3$~THz, which is off-resonant with the FES mode frequency.\\
The existence of an optimal for $\langle \hat{Q}_{\rm f}^2\rangle$ is a consequence between the trade-off between the delocalization and the dipole matrix elements between the phononic states coupled by the cavity photons. Indeed the higher $\omega_{\rm c}$ the higher the phononic excited states that are coupled to the groundstate, which in turns means that resulting delocalization is larger but at the same time the dipole-matrix element becomes 
smaller.\\
Beside the Von Neumann entropy another way to characterize the correlation between light and matter is to evaluate the so-called purity~\cite{flick2015}. This is defined as follows:
\begin{equation}
\gamma={\rm Tr}\left[\hat{\rho}_{\rm ph}^2\right]
\end{equation}
A purity value that deviates from 1 means that the groundstate cannot be factorized in a simple tensor product of a phononic and photonic wavefunction, hence light and matter are correlated.\\
Finally, the results on the expectation value of photon number $N_{\rm ph} = \langle \hat{a}^{\dagger}\hat{a} \rangle$ on the groundstate justifies the use of the term photo-groundstate. Indeed it is clear that even in the cavity is dark, there is a finite number of photons generated by the presence of SrTiO$_3$.

\section{Dynamical Localization induced by Vacuum Fluctuations}
In this section we describe and alternative analytic simple approach to extend the theory of dynamical localization to the case of the quantized light field in a cavity~\cite{Dunlap:1986co,Sentef:2020ho}.
\subsection{Effective Hamiltonian}
The eigenvalue problem associated with the QED Hamiltonian in the first section can be rewritten in the following matrix form:
\begin{equation}
    \left(\begin{array}{cccccc}
         H_0         & H_1           & 0           & 0 & 0 & \ldots \\
         H_1^\dagger & H_0 +\omega & \sqrt{2}H_1 &  0 & 0 & \ldots\\
          0 & \sqrt{2} H_1^\dagger & H_0 +2\omega & \sqrt{3}H_1 & 0 & \ldots\\
          \vdots &  \ddots{} &  \ddots{} & \ddots{} &\vdots & 
          \end{array}
    \right) 
    \left(\begin{array}{c}
         u_0 \\ u_1 \\ u_2 \\ \vdots
         \end{array}
    \right) = 
    E \left(\begin{array}{c}
         u_0 \\ u_1 \\ u_2 \\ \vdots
         \end{array}
    \right)
\end{equation}
where $H_0$ and $H_1$ are matrices in the matter basis (see below) and $u_i$ are photon components of the eigenstate. The action on the $n$th photon-sector can be written as
\begin{equation}
    \sqrt{n} H_1^\dagger u_{n-1} + (H_0 + n\omega)u_n + \sqrt{n+1}H_1u_{n+1} = E u_{n},
\end{equation}
which can be proven by inspection. From this one can recursively write the action of the full Hamiltonian matrix unto a single photon sector in the form $H_{\rm eff}|u_0\rangle = E |u_0\rangle$ and thus get an approximation for the groundstate (or low lying eigenvalues).
The effective Hamiltonian can be easily shown to be:
\begin{equation}
  H_{\rm eff} = H_0 - H_1 \cfrac{1}{H_0 +\omega - E - 2 H_1 \cfrac{1}{H_0 + 2\omega - E+\cdots}H_1^\dagger}H_1^\dagger.
\end{equation}

\subsection{High frequency Approximation}
How many photon sectors need to be included depends on the ratio between light-matter coupling and frequency. To make this clear we write $H_1=A_0P \tilde{H_1}$ and factorize:
\begin{equation}
  H_{\rm eff} = H_0 - \cfrac{(A_0P)^2}{\omega}\tilde{H}_1 \frac{1}{1+ \frac{H_0-E}{\omega} - 2\cfrac{(A_0 P)^2}{\omega} \tilde{H}_1 \frac{1}{H_0 + 2\omega - E}\tilde{H}_1^\dagger}\tilde{H}_1^\dagger
\end{equation}
Only if $\omega >> A_0P$ the continued fractions can be neglected and the leading term reads
\begin{equation}
  H_{\rm eff} = H_0 - \frac{A_0^2P^2}{\omega}\tilde{H}_1 \frac{1}{1+ \frac{H_0-E}{\omega}}\tilde{H}_1^\dagger = H_0 - H_1 \left[H_0 +\omega - E \right]^{-1}H_1^\dagger,
\end{equation}
as in the main text $A_0$ defines the photon mode volume and $P$ is a c-number that sets the scale of the photon-phonon momentum matrix. 
Under this condition we can use the resolvent as a Neumann series and write
\begin{equation}\label{eq:Heff_Neumann}
  H_{\rm eff} = H_0 - H_1 \sum^\infty_{n=0}\frac{(H_0 - E)^n}{\omega^{n+1}}H_1^\dagger .
\end{equation}
The Neumann series converges for $\omega>\max({\{E_{0\lambda} }\})-E$, where $H_0\psi_{\lambda} = E_{0\lambda}\psi_{\lambda}$, but Eq.~(\ref{eq:Heff_Neumann}) is only a valid approximation for the full Hamiltonian as long as the truncation at $n=1$ is possible. 

The series formulation of the effective Hamiltonian is useful because it yield to leading order in $1/\omega$
\begin{equation}
    H^{(1)}_{\rm eff}\approx  H_0 - \frac{H_1 H_1^\dagger}{\omega}   
\end{equation}
which can be readily solved. For all higher orders the effective Hamiltonain gives self-consistent eigenvalue equation that can only be solved iteratively.  However, one can approximate this self-consistency by considering the linearization  $E\rightarrow E_0$. To second order in $1/\omega$ the effective Hamiltonian for the $i$th eigenstate then reads
\begin{equation}
    H^{(2)}_{\rm eff,i}\approx  H_0 - \frac{H_1 H_1^\dagger}{\omega} + \frac{H_1 H_0 H_1^\dagger}{\omega^2} - \frac{H_1 E_{0i} H_1^\dagger}{\omega^2}  
\end{equation}
which has to be solved separately for each eigenstate.

\subsection{Localisation in SrTiO$_3$}
Two describe the photon induced localization we consider the system within a two-level approximation. We choose as the two level two Gaussians which are localized in the left and right well of the 1D FES mode energy potential respectively. In matrix form this translates to:
\begin{equation}
    H =  \left(\begin{array}{cc}
            0 & t\\
            t & 0 \end{array}
         \right) + 
          A_0 P\left(\begin{array}{cc}
            0 & -i \\
            i & 0 \end{array}
         \right) (a+a^\dagger) + \omega a^\dagger a  
\end{equation}
so that $H_0 = t \sigma_x$ and $H_1 = A_0 P \sigma_y$. Here $P$ is directly the L-R momentum matrix element. The high frequency approximation from the previous subsection then reads:
\begin{equation}
    H_{\rm eff} = \left(t  -  \frac{A_0^2 P^2 t }{\omega^2}\right)\sigma_x - \left(\frac{A_0^2 P^2 }{\omega} + \frac{A_0^2 P^2 E_0 }{\omega^2}\right)\mathbf{1} .
\end{equation}
The second term is only shifting the eigenvalues, while the first one gives full localisation if $\frac{A_0^2 P^2 }{\omega^2} = 1$. In that case the effective Hamiltonian has degenerate eigenvalues and similar to the full phonon-QED case one can make linear combinations of eigenvectors that give $(1,0)$ and $(0,1)$.

\section{Temperature Dependent Response Function}
\begin{figure}[t]
  \centering
   \includegraphics[width=\columnwidth]{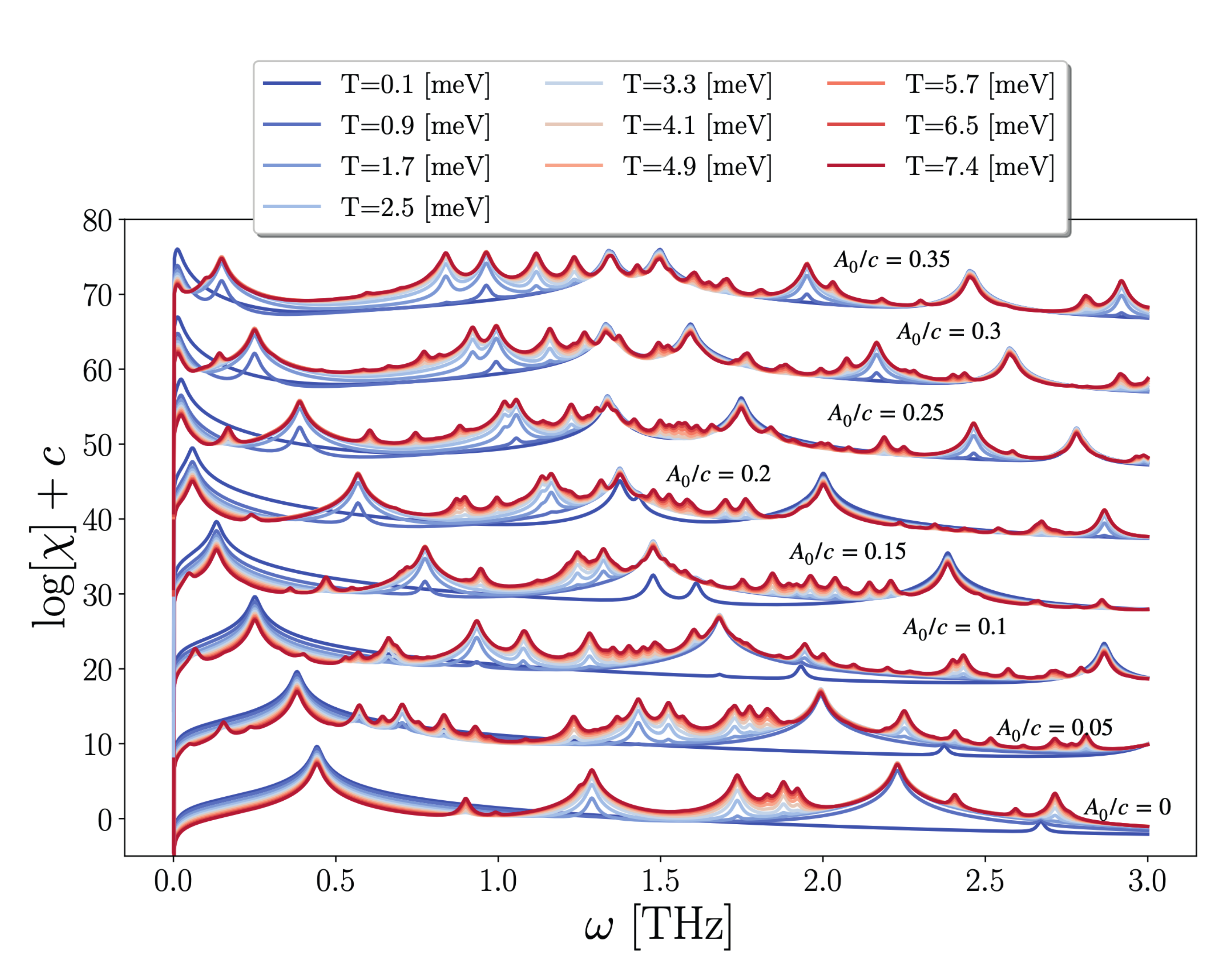}
   \caption{\label{fig:S2}
    Dependence of the imaginary part of the response function on temperature and cavity coupling strenght. The response to a probing electric field is calculated within Kubo's linear response theory. The values are shown in logarithmic scale and they are shifted by a constant for clarity.}
\end{figure}

In order to calculate the phase diagram that we presented in the main text, we need to include the effect of temperature in our theory. To do so we apply Kubo's formula for the linear response of a thermal state to a perturbation described by: 
\begin{equation}
    \hat{H}'(t) = -Z_{\rm f}\hat{Q_{\rm{f}}} E(t)
\end{equation} 
where $Z_{\rm f}$, the FES mode effective charge, is assumed to be temperature independent. 
Applying Kubo's formula, the resulting polarizability takes the form: 
\begin{equation}
\begin{split}
\chi(\omega,T,A_0)=-\sum_{i,j} \rho_i(T, A_0)Z^{*}|D_{ij}(A_0)|^2 &\times \left\{\frac{1}{\left[\epsilon_j(A_0) - \epsilon_i(A_0)\right]-\omega -i\delta }\right. \\
&\left. +\frac{1}{\left[\epsilon_j(A_0) - \epsilon_i(A_0)\right]+\omega +i\delta }\right\},
\end{split}
\end{equation}
where the dipole matrix and the thermal density matrix are defined as $D_{ij}(A_0)=\langle \psi_i(A_0)|\hat{Q}_f|\psi_j(A_0)\rangle$ and $\rho_i(T, A_0)=e^{-[\epsilon_i(A_0)-\epsilon_0(A_0)]/k_BT}/\sum_j e^{-[\epsilon_j(A_0)-\epsilon_0(A_0)]/k_BT}$, respectively and $|\psi_i(A_0)\rangle$ are the eigenstates of the QED Hamiltonian for different values of the cavity coupling.

We then define a characteristic temperature dependent FES mode frequency from such a response function as:
\begin{equation}
    \omega(T, A_0)=\frac{\int d\omega~ \omega~{\rm Im}[\chi(\omega,T A_0)]}{\int d\omega~  {\rm Im}[\chi(\omega,T, A_0)]}.
\end{equation}